\RequirePackage{lineno}
\documentclass[aps,prd,linenumbers,groupedaddress]{revtex4}

\usepackage{graphicx}
\usepackage{bm}
\usepackage{amssymb}
\usepackage{color}
\usepackage{dcolumn}
\usepackage{tabularx}
\usepackage{comment}

\newcolumntype{C}[1]{>
{\hsize=#1\hsize\centering\arraybackslash}X}%

\usepackage{xspace}

\begin{document}

\title{The Coldest Cubic Meter in the Known Universe}
\author{Jonathan~L.~Ouellet}
\affiliation{University of California, Berkeley}
\email{jlouellet@lbl.gov}
\date{\today}

\begin{abstract}
  CUORE is a 741 kg array of TeO$_2$ bolometers that will search for
  the neutrinoless double beta decay of $^{130}$Te. The detector is
  being constructed at the Laboratori Nazionali del Gran Sasso in
  Italy, where it will begin taking data in 2015. The CUORE cryostat
  will cool several metric tonnes of material to below 1~K and the
  CUORE detector itself will operate at a typical temperature of
  10~mK. At this temperature, the CUORE detector will be the coldest
  contiguous cubic meter in the known Universe.
\end{abstract}
\maketitle


\section{Introduction}
The Cryogenic Underground Observatory for Rare Events (CUORE) is an
experiment being built at the Laboratori Nazionali del Gran Sasso. It
designed to search for Neutrinoless Double Beta Decay
(0$\nu\beta\beta$) and is scheduled to begin commissioning in 2015. It
is also going to be the \emph{coldest} cubic meter in the known
Universe.

Neutrinoless double beta decay is a second order weak decay in which a
nucleus spontaneously converts two of its neutrons into protons and
produces two electrons in the process
($(Z,A)\rightarrow(Z+2,A)+2e^-$), without the corresponding electron
anti-neutrinos. If observed, this process would indicate a violation
of lepton number and have major implications for the nature of the
neutrino as well as the fundamental symmetries of the Universe. It
could even shed light on the question of the Baryon asymmetry of the
universe --- the matter/anti-matter mismatch that allowed the
formation of stars and galaxies in the first place.

The CUORE detector will search for the $0\nu\beta\beta$ decay of the
isotope $^{130}$Te using a cryogenic array of TeO$_2$ bolometers. Each
bolometer module consists of a 750~g absorber and an NTD thermistor to
measure temperature. When an individual nucleus anywhere inside the
absorber undergoes a decay, it releases a small amount of energy which
is quickly converted into a rise in temperature of the system, which
is detected by the thermistor. Since the energy deposited is so small,
--- of order a few MeV --- seeing a measurable increase in the
temperature requires a very small heat capacity and thus an operating
temperature near absolute zero. In the case of CUORE, the target
operating temperature is 10~mK and the decays of interest cause
temperature spikes on the order of hundreds of $\mu$K. The full CUORE
detector consists of 988 bolometric modules for a total mass of
741~kg. When operational, this mass, plus an additional 2 metric
tonnes of supporting material, will need to be cooled to 10~mK,
producing the largest region in the Universe at that temperature.

In this paper, I will briefly discuss the temperature of the Universe
and some of the naturally occurring cold places in it. I will
introduce the CUORE detector and cryostat and describe the working
volume and its temperature. I will also compare the CUORE detector to
some of the other large low temperature experiments that are operating
or being built.

\section{Low Temperature Regions in Nature}

In comparison to the CUORE detector, the Universe is actually quite
warm. Its temperature is dominated by the Cosmic Microwave Background
(CMB) photons that pervade all empty space. This thermal bath of
photons exists everywhere throughout the Universe and has a well
defined temperature which has been measured with extreme accuracy to
be $T_{\rm CMB}=2.72548\pm0.00057$~K \cite{Fixsen2009}.

Many regions of space are heated above $T_{\rm CMB}$ by structure
formation and the radiation this gives off, but there is currently
only one known naturally occurring region \emph{below} $T_{\rm CMB}$
and that is the Boomerang Nebula \cite{Wegner1979}. This
proto-planetary nebula (PPN) consists of a central star surrounded by
an envelope of molecular gas. The Boomerang Nebula is unique among
known PPN in that it has produced an extremely massive and rapidly
expanding envelope of gas. The high opacity of this envelope absorbs
CMB photons in the outer layers, shielding the inner regions and
allowing them to cool via adiabatic expansion.  By combining radio
measurements with radiative modeling, the authors of \cite{Sahai1997}
place the kinematic temperature at about 1~K, but possibly as cold as
0.3~K. This makes the Boomerang Nebula the coldest known object in the
Universe \emph{outside the laboratory}. But here on Earth we routinely
achieve temperatures colder than 300~mK.

\section{The CUORE Cryostat}

The CUORE detector is hosted in one of the largest cryostats ever
constructed and is cooled by a $^3$He/$^4$He dilution refrigerator
that was designed and built by Leiden Cryogenics and is one of the
most powerful in the world. A detailed description of the CUORE
cryostat can be found here \cite{Schaeffer2009}, and a paper
describing its commissioning is in preparation. The
cryostat is built as a series of nested vessels that step the
temperature down from 300~K to $\sim40$~K, $\sim3.5$~K, $\sim800$~mK,
$\sim 50$~mK and finally the detector temperature of $10$~mK. Each stage
is connected to a cooling unit and has a radiation shield that
thermally isolates the enclosed volume.

The largest stage of the cryostat below 1~K is the Still. The
temperature of this stage will be adjusted to optimize the temperature
of the coldest stage, but it is typically maintained between about
$\sim$800-900~mK. It is composed of a radiation
shielding copper can 112~cm in diameter by 185~cm in height, mounted
to the bottom of the copper Still plate that has a diameter of 133~cm
and a thickness of 4.3~cm. The total enclosed volume at or below
$\sim1$~K is $\sim$1890~L.

Inside the still shield is the next colder stage of the cryostat, the
Heat Exchanger. The temperature of this stage will also be adjusted to
maintain the base temperature but it is typically maintained at
$\sim$50~mK. This stage consists of a radiation shielding copper can
of diameter of 103~cm and height of 165~cm, it is mounted to the
bottom of a copper plate 107~cm in diameter and 2.8~cm thick. The
total enclosed volume at or below $\sim$50~mK is 1435~L.

Inside the Heat Exchanger (HEx) vessel is the coldest stage of the
cryostat, the Mixing Chamber (MC) plate, the lead shielding, and the
CUORE detector itself. The MC plate is suspended from the Heat
Exchanger plate and hosts the final stage of the dilution
refrigerator. The plate itself is 98~cm in diameter and 1.8~cm thick
and supports a radiation shielding copper vessel below, which is 94~cm in
diameter and 130~cm in height. The MC stage is cooled to 10~mK and
encloses a volume of 990~L.

However, inside this volume things get slightly complicated. Below the
10mK plate, there will be $\sim$2.6 metric tonnes of lead and copper
shielding, which, for reasons of cooling power, is thermalized to
50~mK. This shielding takes up a volume of 235~L inside the 10~mK
shielding. So to be conservative, we will take the operating volume at
10mK to be only that of the shielded detector itself, 636~L. These
sizes and volumes are summarized in Tab. \ref{tab:QStages}.

\begin{table*}[t]
  \caption{The stages of the CUORE cryostat below 1K and the volumes
    and masses colder than that temperature. The values are cumulative
    and should be read as the `total volume/mass colder than.'
    (Numbers are approximate.)
    \cite{CUORECryo1,CUORECryo2,CUORECryo3}}
  \label{tab:QStages}
  \begin{tabularx}{.9\linewidth}{|C{.6}|C{.5}|C{.5}|C{.4}|}
    \hline
    Stage & Temperature (mK)& Volume (L) & Mass (kg)\\
    \hline
    4K Stage & 3.5~K & 3340 & 16000 \\
    Still & 850 & 1890 & 14100\\
    Heat Exchange (HEx)& 50 & 1440 & 5900 \\
    Base (no load) & 10 & 990 & 400\\
    Base & 10 & 636 & 2000\\
    \hline
  \end{tabularx}
\end{table*}
\section{Discussion}

The current record for the coldest cubic meter in the Universe was set
in the first of the CUORE cryostat commissioning runs without the lead
shielding mounted [paper in preparation]. When the CUORE detector is
fully commissioned and running (2015), the detector will be held
stably at the operating temperature of $\sim$10~mK for the duration of
CUORE data taking -- which is expected to be $\sim5$~years. During
this time, both the 636~L held at 10~mK and the 1435~L held entirely
below 50~mK will be the coldest volumes of those respective sizes in
the known Universe. This gives CUORE cryostat the distinction of being
the \emph{Coldest Cubic Meter} in the known Universe.

In Tab. \ref{tab:ColdExperiments}, I list a few, but certainly not
all, of the larger currently running experiments below 100~mK. I list
their approximate cold volumes and operating temperatures. Not
surprisingly, the largest coldest experiments are often rare-event
searches.

\begin{table*}[t]
  \caption{A (non-exhaustive) list of currently running large volume
    experiments with operating temperatures below 100mK. The mass
    represents the \emph{target} or detector mass, excluding any
    supporting material or structure. CUORE (HEx) is not a separate
    experiment, but just the 50~mK stage of the CUORE detector. All
    numbers are approximate, and only meant to give a sense of scale.
    $^a$Auriga/Nautilus operated at $T\sim100$~mK from 1997-99, but
    are currently taking data at 4.4~K.}
  \label{tab:ColdExperiments}
  \footnotesize
  \begin{tabularx}{\linewidth}{|C{.36}|C{.23}|C{.23}|C{.32}|C{.36}|C{.38}|C{.12}|}
    \hline
    Experiment & Mass (kg) & Size (L) & Temperature & Physics Goal & 
    Location & Ref.\\
    \hline
    CUORE & 741 & 636 & 10mK & $0\nu\beta\beta$ & Gran Sasso, Italy &
    \cite{Schaeffer2009}\\
    CUORE-0 & 39 & 27 & 12mK & $0\nu\beta\beta$ & Gran Sasso, Italy & \\
    CRESST-II & 10 & 24 & 15mK & Dark Matter & Gran Sasso, Italy & 
    \cite{Lang2008}\\
    Edelweiss & 32 & 50 & 20mK & Dark Matter & Modane, France&
    \cite{EDELWEISSsite} \\
    SuperCDMS & 10 & 21 & 40mK & Dark Matter & Soudan, SD & 
    \cite{CDMSCollaboration1999}\\
    CUORE (HEx) & - & 1435 & 50mK & $0\nu\beta\beta$ & Gran Sasso, Italy & 
    \cite{Schaeffer2009}\\
    Auriga/Nautilus$^{a}$ & 2200 & 848L & 100mK & Gravity Wave & Italy & \\
    \hline
  \end{tabularx}
\end{table*}

\section{Caveats, Qualifications, Ifs and Buts..}
There are several caveats and assumptions that should be noted. Many
of these are technicalities, but they deserve mentioning:

\begin{itemize}
\item The CUORE detector, like everything on Earth, is bathed in a
  constant flux of neutrinos both from the sun and earth's core. The
  solar neutrinos were last thermalized in the core of the sun to
  temperatures of order $10^7$~K and neutrinos from the earth's core
  were last thermalized to temperatures of order $10^4$~K. However,
  neutrinos interact so infrequently --- we expect of order 100 solar
  neutrinos per year to interact in the CUORE detector --- that they
  never reach a thermal equilibrium with anything on Earth
  (thankfully). However, all of these neutrinos are still technically
  present inside the volume of the CUORE detector, so I explicitly
  ignore them here.

\item Another very interesting source of neutrinos that are also
  present inside the volume of the CUORE detector is the Cosmic
  Neutrino Background (C$\nu$B). These are the relic neutrinos from
  the Big Bang, and like the photons in the CMB they are expected to
  pervade the entire Universe. Unlike the photons in the CMB, these
  neutrinos interact so infrequently that they have not actually been
  detected yet. These neutrinos are expected to be warmer than the
  CUORE detector but like the solar and geo-neutrinos, never come into
  thermal equilibrium, so I explicitly ignore these as well.

\item CUORE will have a significant amount of lead shielding sitting
  inside the 10~mK shield and thermalized to 50~mK. Lead becomes
  superconducting around 7~K and as a result, as the temperature drops
  below $\sim T_c/10=700$~mK, the thermal conductivity becomes very
  poor and the lead begins to self insulate. It will eventually reach
  50~mK, but it is difficult to say on what timescale. Neglecting the
  lead in the 10~mK volume, the remaining volume at 50~mK is about
  1245~L.

\item The definition of the cubic meter can be made fuzzy. Here
  I have considered only a simple contiguous convex volume of space. 

\item Proving the non-existence of a cubic meter in the Universe
  colder than CUORE is, of course, an impossible task. I have
  implicitly restricted the discussion to known or discovered
  phenomena. But, I admit the possibility of another planet somewhere
  in an infinite Universe, which is entirely identical to Earth in
  every way, except that their CUORE collaboration has decided to
  operate their CUORE detector at 9~mK.

\end{itemize}

\section{Conclusion}

The volume inside the CUORE detector will be the largest volume in the
Universe below 50~mK and the largest volume at 10~mK. It has already
achieved this temperature in its first test cooldown and can now be
called the coldest cubic meter in the known Universe. And as such, the
CUORE detector has a strong claim to the title of the coldest
\emph{place} in the Universe as well. 

\section*{Acknowledgements}

I would like to thank the entire CUORE collaboration, of which I am
only a relatively recent member. Producing the coldest cubic meter in
the Universe is the result of years of dedication and I am grateful to
be a part of such an ambitious enterprise. In particular, I would like
to thank Francesco Terranova for the detailed accounting of the
cryostat volume, Luca Taffarello for the most recent information on
AURIGA and Nautilus, and Yury Kolomensky for the very useful
discussions. The CUORE collaboration consists of institutions in
Italy, the United States, Spain, China and France. This work was
supported by the National Science Foundation under grant
Nos. PHY-0902171 and PHY-1314881. This material is also based upon
work supported by the US Department of Energy (DOE) Office of Science
under Contract Nos. DE-AC02-05CH11231 and DE-AC52-07NA27344; and by
the DOE Office of Science, Office of Nuclear Physics under Contract
Nos. DE-FG02-08ER41551 and DEFG03-00ER41138.

\bibliography{biblio}

\appendix
\section{More on the CMB}

The CMB sets the standard for what we consider the temperature of the
Universe, and in this appendix we consider the possibility of a purely
\emph{statistical} fluctuation of the CMB temperature down to 10~mK
somewhere in the Universe. If we broaden our working definition of
temperature, we can calculate the probability that the \emph{mean}
kinetic energy of all the CMB photons in a single cubic meter of space
fluctuates down to the point that their mean energy is consistent with
$\sim10\text{~mK}$. (I should warn, that the following argument will be
\emph{very} approximate.)

Starting from the Planck distribution
\[
n(E)dE=\frac{1}{\pi^2\hbar^3c^3}\frac{E^2}{e^{E/k_BT}-1}dE
\]
we can integrate this to find the photon number density at 2.7~K,
$n\sim4\times10^8\text{~m}^{-3}$, the mean photon energy $\langle
E\rangle\approx0.7\text{~meV}$, and the RMS $\sigma_E\sim0.4$~meV. Due
to the extremely large number of photons in a given cubic meter, the
magnitude of fluctuations of \emph{mean} kinetic energy is extremely
suppressed, $\sigma_{\langle
  E\rangle}\sim\frac{\sigma_E}{\sqrt{\langle N \rangle}}\sim$20~neV. A
fluctuation down to a mean kinetic energy consistent with $T\sim
10$~mK (or $\langle E\rangle\sim2.3~\mu$eV), amounts to a
$\sim31,000\sigma$ downward fluctuation. 

If one were to assume a Gaussian probability distribution, this
calculation would imply that the fraction of cubic meters in the
Universe that had fluctuated down this far would be about 1 in
$\sim10^{10^8}$. For reference, there are $\sim10^{80}$ cubic meters
in the observable Universe. However, we are extrapolating this
distribution down so far that considering it Gaussian is certainly no
longer warranted. 


So instead we ask the question what \emph{is} the coldest cubic meter
in the CMB --- according to this interpretation of mean energy
density. Or in other words, what temperature corresponds to the mean
energy density which we expect fewer than 1 in $10^{80}$ cubic meters
of space to be colder than. Assuming a Gaussian distribution, this
would take us down only 19$\sigma$ --- a much more modest distance to
extrapolate than 31,000$\sigma$. This corresponds to a fractional
difference in temperature of $\frac{\Delta
  T}{T}\sim6.1\times10^{-4}$. Or in other words, for statistical
fluctuations alone, we expect that at any given time the coldest cubic
meter anywhere in the CMB is only $\sim1.6$~mK colder than the average
temperature of the CMB.

\end{document}